\documentclass{ws-mpla}

\usepackage{graphicx, color}
\usepackage{dcolumn}
\usepackage{bm}

\usepackage{latexsym}
\usepackage{mathrsfs}
\usepackage{amssymb}

\usepackage[hyperfigures=false, colorlinks=true, pagecolor=black,
linkcolor=black, citecolor=black, urlcolor=black, pagebackref=false,
hyperfigures=false, bookmarks=true, bookmarksopen=false, pdfauthor={Ashkbiz Danehkar}, pdftitle={On the Significance of the Weyl Curvature in a Relativistic Cosmological Model},
pdfsubject={PACS numbers: 98.80.-k, 98.80.Jk, 47.75.+f}, pdfkeywords={Relativistic cosmology; Weyl curvature; covariant formalism}]{hyperref}%

\usepackage[numbers,sort&compress,super]{natbib}

\newcommand{\onlinecite}[1]{\hspace{-1 ex} \nocite{#1}\citenum{#1}} 

\begin{document}

\markboth{A. Danehkar} {On the Significance of the Weyl Curvature}

\catchline{}{}{}{}{}

\title{ON THE SIGNIFICANCE OF THE WEYL CURVATURE\\ IN A RELATIVISTIC COSMOLOGICAL MODEL}

\author{ASHKBIZ DANEHKAR\footnote{Present Address: School of Mathematics and Physics, Queen’s University, Belfast BT7 1NN, UK. E-mail: adanehkar01@qub.ac.uk}}

\address{Faculty of Physics, University of Craiova, 13~Al.\,I.\,Cuza Str., 200585 Craiova, Romania\\
danehkar@central.ucv.ro}

\maketitle

\pub{Received 15 March 2008}{Revised 5 August 2009}

\begin{abstract}
The Weyl curvature includes the Newtonian field and an additional
field, the so-called anti-Newtonian. In this paper, we use the
Bianchi and Ricci identities to provide a set of constraints and
propagations for the Weyl fields. The temporal evolutions of
propagations manifest explicit solutions of gravitational waves. We
see that models with purely Newtonian field are inconsistent with
relativistic models and obstruct sounding solutions. Therefore, both
fields are necessary for the nonlocal nature and radiative solutions
of gravitation. 
\end{abstract}

\ccode{PACS Nos.: 98.80.-k, 98.80.Jk, 47.75.+f}

\section{Introduction}

\label{intro}

In the theory of general relativity, one can split the Riemann
curvature tensor into the Ricci tensor defined by the Einstein
equation and the Weyl curvature
tensor.\cite{Jordean1960,Ciufolini1995,Hawking1973,Misner1973} 
Additionally, one can split the Weyl tensor into the electric part
and the
magnetic part, the so-called gravitoelectric/-magnetic fields,\cite{Thorne1980} being due to some similarity to electrodynamical
counterparts.\cite{Ciufolini1995,Pirani1957,Pirani1962a,Pirani1962b,Kofman1995} 
We describe the gravitoelectric field as the tidal (Newtonian)
force,\cite{Kofman1995,Ellis1973} but the gravitomagnetic field has
no Newtonian analogy, called anti-Newtonian. Nonlocal
characteristics arising from the Weyl curvature provides a
description of the Newtonian force, although the Einstein equation
describes a local dynamics of
spacetime.\cite{Kofman1995,Wiltshire2008} The Weyl curvature also
includes an additional force: the gravitomagnetic field that is
produced by the mass currents analogously an electric current
generating a magnetic field.\cite{Ciufolini1995} In fact, the theory
of general relativity predicts two main concepts: gravitomagnetic
fields and gravitational waves. Gravitation similar to
electromagnetism propagates at identical speed, that provides a
sounding analysis and a radiative description of force. We notice
the Weyl tensor encoding the tidal force, a new force by its
magnetic part, and a treatment of gravitational waves.

Determination of gravitational waves and gravitomagnetism (new
force) is experimental tests of general relativity.\cite{Will2006} 
Gravitational radiation of a binary system of compact objects has
been proposed to be detected by a resonant bar\cite{Hamilton1992} or
a laser interferometer in space,\cite{Bender1989,Faller1989} such as
the LIGO\cite{Fritschel1998} and VIRGO.\cite{Brillet1998} A
non-rotating compact object produces the standard Schwarzschild
field, whereas a rotating body also generates the gravitomagnetic
field. It has been suggested as a mechanism for the jet formation in
quasars and galactic nuclei.\cite{Bardeen1975,Thorne1986} The
resulting action of the gravitomagnetic fields and of the viscous
forces implies that the formation of the accretion disk into the
equatorial plane of the central body while the jets are ejected
along angular momentum vector perpendicularly to the equatorial
plane.\cite{Ciufolini1995,Bardeen1975} The gravitomagnetic field
implies that a rotating body e.g. the Earth affects the
motion of orbiting satellites. This effect has been recently
measured using the LAGEOS I and LAGEOS II
satellites.\cite{Ciufolini2002} However, we may need counting some
possible errors in the LAGEOS data.\cite{Iorio2005} Using two recent
orbiting geodesy satellites (CHAMP and GRACE), it has been reported
confirmation of general relativity with a total error between $5\%$
and $10\%$.\cite{Ciufolini2004,Ciufolini2006,Ciufolini2008}

In this paper, we describe kinematic and dynamic equations of the
Weyl curvature variables, i.\thinspace e., the gravitoelectric field
as the relativistic generalization of the tidal forces and the
gravitomagnetic field in a cosmological model containing the
relativistic fluid description of matter. We use the convention
based on $8\pi G=1=c$. We denote the round brackets enclosing
indices for symmetrization, and the square brackets for
antisymmetrization. The organization of this paper is as follows. In
Sec.~\ref{sec2}, we introduce the 3 + 1 covariant formalism,
kinematic quantities, and dynamic quantities in a hydrodynamic
description of matter. In Sec.~\ref{sec3}, we obtain constraint and
propagation equations for the Weyl fields from the Bianchi and Ricci
identities. In Sec.~\ref{sec4}, rotation and distortion are
characterized as wave solutions. In Sec.~\ref{sec5}, we study a
Newtonian model as purely gravitoelectric in an irrotational static
spacetime and a perfect-fluid model, and an anti-Newtonian model as
purely gravitomagnetic in a shearless static and perfect-fluid
model. We see that both models are generally inconsistent with
relativistic models, allowing no possibility for wave solutions.
Section~\ref{sec6} provides a conclusion.

\section{Covariant Formalism}

\label{sec2}

According to the pattern of classical hydrodynamics, we decompose
the spacetime metric into the spatial metric and the instantaneous
rest-space of a comoving observer. The formalism, known as the 3 + 1
covariant approach to general
relativity,\cite{Raychaudhuri1955,Raychaudhuri1957,Raychaudhuri1979,Heckmann1955,Heckmann1956,Heckmann1959} 
has been used for numerous
applications.\cite{Ellis1973,Ehlers1993a,Ehlers1993b,Ellis1971,Ellis1989,King1973} 
In this approach, we rewrite equations governing relativistic fluid
dynamics by using projected vectors and projected symmetric
traceless tensors instead of metrics.\cite{Ellis1973,Ellis1989}

We take a four-velocity vector $u^{a}$ field in a given (3 + 1)-dimensional 
spacetime to be a unit vector field $u^{a}u_{a}=-1$. We
define a spatial metric (or projector tensor)
$h_{ab}=g_{ab}+u_{a}u_{b}$, where $g_{ab}$ is the spacetime metric.
It decomposes the spacetime metric into the spatial metric and the
instantaneous rest-space of an observer moving with four-velocity
$u^{a}$.\cite{Ciufolini1995,Ellis1971,Cattaneo1959} We get some
properties for the spatial metric
\begin{equation}%
\begin{array}
[c]{ccc}%
{h_{ab}u^{b}=0,} & {~~~~~~~~h_{a}{}^{c}h_{cb}=h_{ab},} & {~~~~~~~~h_{a}{}^{a}=3.}%
\end{array}
\label{eq:e_3_3}%
\end{equation}
We also define the spatial alternating tensor as
\begin{equation}
\varepsilon_{abc}=\eta_{abcd}u^{d}, \label{eq:e_3_4}%
\end{equation}
where ${\eta_{abcd}}${\ is }the spacetime alternating tensor,
\begin{equation}%
\begin{array}
[c]{ccc}%
{\eta_{abcd}=-4!\sqrt{|g|}\delta^{0}{}_{[a}\delta^{1}{}_{b}\delta^{2}{}%
_{c}\delta^{3}{}_{d]},} & {~~~~~~~~\delta_{a}{}^{b}=g_{ac}g^{cb},} &
{~~~~~~~~|g|=\det
g_{ab}.}%
\end{array}
\label{eq:e_3_4_1}%
\end{equation}

The covariant spacetime derivative $\nabla_{a}$ is split into a
covariant temporal derivative
\begin{equation}
\dot{T}_{a\cdots}=u^{b}\nabla_{b}T_{a\cdots}, \label{eq:e_3_6}%
\end{equation}
and a covariant spatial derivative
\begin{equation}
\mathrm{D}_{b}T_{a\cdots}=h_{b}{}^{d}h_{a}{}^{c}\cdots\nabla_{d}T_{c\cdots}.
\label{eq:e_3_7}%
\end{equation}
The projected vectors and the projected symmetric traceless parts of
rank-2 tensors are defined by
\begin{equation}%
\begin{array}
[c]{cc}%
{V_{\left\langle a\right\rangle }\equiv h_{a}{}^{b}V_{b},} &
{S_{\left\langle {ab}\right\rangle }\equiv\left\{
{h_{(a}{}^{c}h_{b)}{}^{d}-{\textstyle{\frac
{1}{3}}}h^{cd}h_{ab}}\right\}  S_{cd}.}%
\end{array}
\label{eq:e_3_5}%
\end{equation}
The equations governing these quantities involve a vector product
and its generalization to rank-2 tensors:
\begin{equation}%
\begin{array}
[c]{cc}%
{[V,W]_{a}\equiv\varepsilon_{abc}V^{b}W^{c},} & {~~~~~~~~[S,Q]_{a}\equiv
\varepsilon_{abc}S^{b}{}_{d}Q^{cd},}%
\end{array}
\label{eq:e_3_8}%
\end{equation}%
\begin{equation}%
\begin{array}
[c]{cc}%
{[V,S]_{ab}\equiv\varepsilon_{cd(a}S_{b)}{}^{c}V^{d},} &
{~~~~~~~~[V,S]_{\left\langle {ab}\right\rangle
}\equiv\varepsilon_{cd\left\langle a\right.  }S_{\left.
b\right\rangle }{}^{c}V^{d}.}%
\end{array}
\label{eq:e_3_9}%
\end{equation}
We define divergences and rotations as
\begin{equation}%
\begin{array}
[c]{cc}%
{\mathrm{div}(V)\equiv\mathrm{D}^{a}V_{a},} & {~~~~~~~~(\mathrm{div}S)_{a}%
\equiv\mathrm{D}^{b}S_{ab},}%
\end{array}
\label{eq:e_3_10}%
\end{equation}%
\begin{equation}%
\begin{array}
[c]{cc}%
({\mathrm{curl}V)_{a}\equiv\varepsilon_{abc}\mathrm{D}^{b}V^{c},} &
{~~~~~(\mathrm{curl}S)_{ab}\equiv\varepsilon_{cd(a}\mathrm{D}^{c}S_{b)}{}^{d},}
\\ \\
{(\mathrm{curl}S)_{\left\langle {ab}\right\rangle
}\equiv\varepsilon
_{cd\left\langle a\right.  }\mathrm{D}^{c}S_{\left.  b\right\rangle }{}^{d}.}&{}
\end{array}
\label{eq:e_3_11}%
\end{equation}
We know that
$\mathrm{D}_{c}h_{ab}=0=\mathrm{D}_{d}\varepsilon_{abc}$,
$\dot{h}_{ab}=2u_{(a}\dot{u}_{b)}$, and $\dot{\varepsilon}_{abc}%
=3u_{[a}\varepsilon_{bc]d}\dot{u}^{d}$, then
$u^{a}\dot{h}_{ab}=-\dot{u}_{b}$ and
$u^{a}\dot{\varepsilon}_{abc}=-\dot{u}^{a}\varepsilon_{abc}$. From
these points one can also define\ the relativistically temporal
rotations as
\begin{equation}%
\begin{array}
[c]{ccc}%
{[\dot{u},V]_{a}=-u^{c}\dot{\varepsilon}_{abc}V^{b},} & {~~~~~~~~[\dot{u}%
,S]_{ab}=-u^{c}\dot{\varepsilon}_{cd(a}S_{b)}{}^{d},} \\ \\
{[\dot{u}%
,S]_{\left\langle {ab}\right\rangle
}=-u^{c}\dot{\varepsilon}_{cd\left\langle
a\right.  }S_{\left.  b\right\rangle }{}^{d}.} & {} 
\end{array}
\label{eq:e_3_12}%
\end{equation}
The covariant spatial distortions are
\begin{equation}
\mathrm{D}_{\left\langle a\right.  }V_{\left.  b\right\rangle }=\mathrm{D}%
_{(a}V_{b)}-{\textstyle{\frac{1}{3}}}(\mathrm{div}V)h_{ab}, \label{eq:e_3_13}%
\end{equation}%
\begin{equation}
\mathrm{D}_{\left\langle a\right.  }S_{\left.  {bc}\right\rangle }%
=\mathrm{D}_{(a}S_{bc)}-{\textstyle{\frac{2}{5}}}h_{(ab}(\mathrm{div}S)_{c)}.
\label{eq:e_3_14}%
\end{equation}
We decompose the covariant derivatives of scalars, vectors, and
rank-2 tensors into irreducible components
\begin{equation}
\nabla_{a}f=-\dot{f}u_{a}+\mathrm{D}_{a}f, \label{eq:e_3_22}%
\end{equation}%
\begin{equation}
\nabla_{b}V_{a}=-\left(  {\dot{V}_{\left\langle a\right\rangle }u_{b}%
+u_{a}u_{b}\dot{u}_{c}V^{c}-{\textstyle{\frac{1}{3}}}\Theta u_{a}V_{b}%
-u_{a}\sigma_{bc}V^{c}-u_{a}[\omega,V]_{b}}\right)
+\mathrm{D}_{a}V_{b},
\label{eq:e_3_23}%
\end{equation}%
\begin{align}
\nabla_{c}S_{ab}=&-\Big({\dot{S}_{\left\langle {ab}\right\rangle }u_{c}%
+2u_{(a}S_{b)d}\dot{u}^{d}u_{c}-{\textstyle{\frac{2}{3}}}\Theta u_{(a}S_{b)c}%
}-2u_{(a}S_{b)}{}^{d}\sigma_{dc}~   \nonumber\\
& {-2\varepsilon_{cde}u_{(a}S_{b)}{}^{d}\omega^{e}}\Big)    +\mathrm{D}%
_{a}S_{bc}, \label{eq:e_3_24}%
\end{align}
where
\begin{equation}
\mathrm{D}_{a}V_{b}={\textstyle{\frac{1}{3}}}\mathrm{D}_{c}V^{c}%
h_{ab}-{\textstyle{\frac{1}{2}}}\varepsilon_{abc}\mathrm{curl}V^{c}%
+\mathrm{D}_{\left\langle a\right.  }V_{\left.  b\right\rangle },
\label{eq:e_3_25}%
\end{equation}%
\begin{equation}
\mathrm{D}_{a}S_{bc}={\textstyle{\frac{3}{5}}}\mathrm{D}^{d}S_{d\left\langle
a\right.  }h_{\left.  b\right\rangle c}-{\textstyle{\frac{2}{3}}}%
\varepsilon_{dc(a}\mathrm{curl}S_{b)}{}^{d}+\mathrm{D}_{\left\langle
a\right.
}S_{\left.  {bc}\right\rangle }. \label{eq:e_3_26}%
\end{equation}

We also introduce the kinematic quantities encoding the relative
motion of fluids:
\begin{equation}
\nabla_{b}u_{a}=\mathrm{D}_{b}u_{a}-\dot{u}_{a}u_{b}, \label{eq:e_3_15}%
\end{equation}%
\begin{equation}
\mathrm{D}_{b}u_{a}={\textstyle{\frac{1}{3}}}\Theta h_{ab}+\sigma_{ab}%
+\omega_{ab}, \label{eq:e_3_16}%
\end{equation}
where $\dot{u}_{a}=u^{b}\nabla_{b}u_{a}$ is the relativistic
acceleration
vector, in the frames of instantaneously comoving observers $\dot{u}_{a}%
=\dot{u}_{\left\langle a\right\rangle }$,
$\Theta=\mathrm{D}^{a}u_{a}$ the rate of expansion of fluids,
$\sigma_{ab}=\mathrm{D}_{\left\langle a\right.
}u_{\left.  b\right\rangle }=\mathrm{D}_{(a}u_{b)}-{\textstyle{\frac{1}{3}}%
}h_{ab}\mathrm{D}_{c}u^{c}$ a traceless symmetric tensor ($\sigma_{ab}%
=\sigma_{(ab)}$, $\sigma_{a}{}^{a}=0$); the shear tensor describing
the rate of distortion of fluids, and
$\omega_{ab}=\mathrm{D}_{[a}u_{b]}$ a
skew-symmetric tensor ($\omega_{ab}=\omega_{\lbrack ab]}$, $\omega_{a}{}%
^{a}=0$); the vorticity tensor describing the rotation of
fluids.\cite{Raychaudhuri1979,Ellis1971,Ryan1975}

The vorticity vector\cite{Godel1949,Godel1950} $\omega_{a}$ is
defined by
\begin{equation}
\omega_{a}=-{\textstyle{\frac{1}{2}}}\varepsilon_{abc}\omega^{bc},
\label{eq:e_3_17}%
\end{equation}
where $\omega_{a}u^{a}=0$, $\omega_{ab}\omega^{b}=0$ and the
magnitude
$\omega^{2}={\textstyle{\frac{1}{2}}}\omega_{ab}\omega^{ab}\geq0$ have been imposed.
Accordingly, we obtain
\begin{equation}
\omega_{a}=-{\textstyle{\frac{1}{2}}}\varepsilon_{abc}\mathrm{D}^{b}u^{c}.
\label{eq:e_3_18}%
\end{equation}
The sign convention is such that in the Newtonian theory
$\vec{\omega }=-{\textstyle{\frac{1}{2}}}\vec{\nabla}\times\vec{u}$.

We denote the covariant shear and vorticity products of the
symmetric traceless tensors as
\begin{equation}%
\begin{array}
[c]{cc}%
{[\sigma,S]_{a} = \varepsilon_{abc} \sigma^{b} {}_{d} S^{cd} ,} &
{[\omega,S]_{\left\langle {ab} \right\rangle } =
\varepsilon_{cd\left\langle a
\right.  } S_{\left.  b \right\rangle } {}^{c} \omega^{d} .}%
\end{array}
\label{eq:e_3_21}%
\end{equation}

The energy density and pressure of fluids are encoded in the dynamic
quantities, which generally have the contributions from the energy
flux and anisotropic pressure:
\begin{equation}
T_{ab}=\rho u_{a}u_{b}+ph_{ab}+2q_{(a}u_{b)}+\pi_{ab}, \label{eq:e_3_19}%
\end{equation}%
\begin{equation}%
\begin{array}
[c]{cccc}%
{q_{a}u^{a}=0,} & {~~~~~~~~\pi^{a}{}_{a}=0,} & {~~~~~~~~\pi_{ab}=\pi_{(ab)},} & {~~~~~~~~\pi_{ab}%
u^{b}=0,}%
\end{array}
\label{eq:e_3_20}%
\end{equation}
where $\rho=T_{ab}u^{a}u^{b}$ is the relativistic energy density
relative to $u^{a}$, $p={\textstyle{\frac{1}{3}}}T_{ab}h^{ab}$ the
pressure, $q_{a}=-T_{\left\langle a\right\rangle
b}u^{b}=-h_{a}{}^{c}T_{cb}u^{b}$ the
energy flux relative to $u^{a}$, and $\pi_{ab}=T_{\left\langle {ab}%
\right\rangle }=T_{cd}h^{c}{}_{\left\langle a\right.
}u^{d}{}_{\left.
b\right\rangle }=\left(  {h^{c}{}_{(a}u^{d}{}_{b)}-{\textstyle{\frac{1}{3}}%
}h_{ab}h^{cd}}\right)  T_{cd}$ the traceless anisotropic stress.
Imposing $q^{a}=\pi_{ab}=0$, we get the solution of a perfect fluid
with $T_{ab}=\rho u_{a}u_{b}+ph_{ab}$. In addition $p=0$ gives the
pressure-free matter or dust
solution.\cite{Raychaudhuri1979,Ellis1971,Ryan1975}

\section{Cosmological Field Equations}

\label{sec3}

In the theory of general relativity, we describe the local nature of
gravitational field nearby matter as an algebraic relation between
the Ricci curvature and the matter fields, i.\thinspace e., the
Einstein field equations:
\begin{equation}
R_{ab}=T_{ab}-{\textstyle{\frac{1}{2}}}Tg_{ab}, \label{eq:e_2_1}%
\end{equation}
where $R_{ab}$ is the Ricci curvature, $T_{ab}$ the energy--momentum
of the matter fields, and $T=T_{c}{}^{c}$ the trace of the
energy--momentum tensor.

The successive contractions of Eq. (\ref{eq:e_2_1}) on using of Eq.
(\ref{eq:e_3_19}) lead to a set of relations:
\begin{equation}%
\begin{array}
[c]{cc}%
{R_{ab}u^{a}u^{b}={\textstyle{\frac{1}{2}}}(\rho+3p),} & {~~~~~~~~h_{a}{}^{b}%
R_{bc}u^{c}=-q_{a},} \\ \\
{h_{a}{}^{c}h_{b}{}^{d}R_{cd}={\textstyle{\frac{1}{2}}%
}(\rho-p)h_{ab}+\pi_{ab},} & { }
\end{array}
\label{eq:e_2_2}%
\end{equation}%
\begin{equation}%
\begin{array}
[c]{ccc}%
{R=R_{a}{}^{a},} & {~~~~~~~~T=T_{a}{}^{a}=-\rho+3p,} & {~~~~~~~~R=-T,}%
\end{array}
\label{eq:e_2_3}%
\end{equation}
where $R$ is the Ricci scalar. The Ricci curvature is derived from
the once contracted Riemann curvature tensor:
$R_{ab}=R^{c}{}_{acb}$.

The Riemann tensor is split into symmetric (massless) traceless
$C_{abcd}$ and traceful massive $M_{abcd}$ parts:
\begin{equation}
R_{abcd}=C_{abcd}+M_{abcd}. \label{eq:e_2_4}%
\end{equation}
The symmetric traceless part of the Riemann curvature is called the
Weyl conformal curvature with the following properties:
\begin{equation}%
\begin{array}
[c]{cc}%
{C_{abcd}=C_{[ab][cd]},} & {~~~~~~~~C^{a}{}_{bca}=0=C_{a[bcd]}.}%
\end{array}
\label{eq:e_2_5}%
\end{equation}
The nonlocal (long-range) fields, the parts of the curvature not
directly determined locally by matter, are given by the Weyl
curvature; propagating the Newtonian (and anti-Newtonian) forces and
gravitational waves. It can be shown that the Weyl tensor $C_{abcd}$
is irreducibly split into the Newtonian $C_{abcd}^{\mathrm{N}}$ and
the anti-Newtonian $C_{abcd}^{\mathrm{AN}}$ parts:
\begin{equation}
C_{abcd}=C_{abcd}^{\mathrm{N}}+C_{abcd}^{\mathrm{AN}}, \label{eq:e_2_6}%
\end{equation}%
\begin{equation}
C_{\mathrm{N}}^{ab}{}_{cd}=4\{u^{[a}u_{[c}+h^{[a}{}_{[c}\}E^{b]}{}_{d]},
\label{eq:e_2_7}%
\end{equation}%
\begin{equation}
C_{abcd}^{\mathrm{AN}}=2\varepsilon_{abe}u_{[c}H_{d]}{}^{e}+2\varepsilon
_{cde}u_{[a}H_{b]}{}^{e}, \label{eq:e_2_8}%
\end{equation}
where $E_{ab}=C_{acbd}u^{c}u^{d}$ is the gravitoelectric field and
$H_{ab}={\textstyle{\frac{1}{2}}}\varepsilon_{acd}C^{cd}{}_{be}u^{e}$
the gravitomagnetic field. They are spacelike and traceless
symmetric.

The traceful massive part of the Riemann curvature consists of the
matter fields and the characteristics of local interactions with
matter
\begin{align}
M^{ab}{}_{cd}  &  ={\textstyle{\frac{2}{3}}}(\rho+3p)u^{[a}u_{[c}h^{b]}{}%
_{d]}+{\textstyle{\frac{2}{3}}}\rho h^{a}{}_{[c}h^{b}{}_{d]}\nonumber\\
&
-2u^{[a}h^{b]}{}_{[c}q_{d]}-2u_{[c}h^{[a}{}_{d]}q^{b]}-2u^{[a}u_{[c}\pi
{}^{b]}{}_{d]}+2h^{[a}{}_{[c}\pi^{b]}{}_{d]}, \label{eq:e_2_9}%
\end{align}
Therefore, the Weyl curvature is linked to the matter fields through
the Riemann curvature.

\subsection{Dynamic Formulas}

To provide equations governing relativistic dynamics of matter, we
use the \textit{Bianchi identities}
\begin{equation}
\nabla_{\lbrack e}R_{ab]cd}=0. \label{eq:e_2_10}%
\end{equation}
On substituting Eq. (\ref{eq:e_2_4}) into Eq. (\ref{eq:e_2_10}), we
get the dynamic formula for the Weyl conformal
curvature\cite{Hawking1973,Kundt1960,Kundt1962}:
\begin{equation}
\nabla^{d}C_{abcd}=-\nabla_{\lbrack a}(R_{b]c}-{\textstyle{\frac{1}{6}}%
}g_{b]c}R)=-\nabla_{\lbrack a}(T_{b]c}-{\textstyle{\frac{1}{3}}}g_{b]c}T_{d}%
{}^{d})\equiv J_{abc}. \label{eq:e_2_11}%
\end{equation}
On decomposing Eq. (\ref{eq:e_2_11}) along and orthogonal to a
4-velocity
vector, we obtain constraint ($C^{1,2}{}_{a}$) and propagation ($P^{1,2}%
{}_{ab}$) equations of the Weyl fields in a form analogous to the
Maxwell
equations\cite{Ellis1973,Trumper1964,Trumper1965,Trumper1967}:
\begin{align}
C^{1}{}_{a}\equiv & {(\mathrm{div}E)_{a}}-3\omega^{b}H_{ab}-[\sigma
,H]_{a}-{\textstyle{\frac{1}{3}}}\mathrm{D}_{a}\rho+{\textstyle{\frac{1}{3}}%
}\Theta q_{a}   \nonumber\\
& -{\textstyle{\frac{1}{2}}}\sigma_{ab}q^{b}+{\textstyle{\frac{3}{2}}}%
[\omega,q]_{a}+{\textstyle{\frac{1}{2}}(\mathrm{div}\pi)_{a}}  =0,
\label{eq:e_2_12}%
\end{align}%
\begin{align}
C^{2}{}_{a}\equiv & {(\mathrm{div}H)_{a}}+3\omega^{b}E_{ab}+[\sigma,E]_{a}%
+\omega_{a}(\rho+p)  & \nonumber\\
& +{\textstyle{\frac{1}{2}}}\mathrm{curl}(q)_{a}+{\textstyle{\frac{1}{2}}%
}[\sigma,\pi]_{a}-{\textstyle{\frac{1}{2}}}\omega^{b}\pi_{ab}   =0,
\label{eq:e_2_13}%
\end{align}%
\begin{align}
P^{1}{}_{ab}\equiv & \mathrm{curl}(H)_{ab}+2[\dot{u},H]_{\left\langle
{ab}\right\rangle }-\dot{E}_{\left\langle {ab}\right\rangle }-\Theta
E_{ab}+[\omega,E]_{\left\langle {ab}\right\rangle }  \nonumber\\
& +3\sigma_{c\left\langle a\right.  }E_{\left.  b\right\rangle }{}%
^{c}-{\textstyle{\frac{1}{2}}}\sigma_{ab}(\rho+p)-{\textstyle{\frac{1}{2}}%
}\mathrm{D}_{\left\langle a\right.  }q_{\left.  b\right\rangle
}-\dot
{u}_{\left\langle a\right.  }q_{\left.  b\right\rangle }  \nonumber\\
& -{\textstyle{\frac{1}{2}}}\dot{\pi}_{\left\langle {ab}\right\rangle
}-{\textstyle{\frac{1}{6}}}\Theta\pi_{ab}+{\textstyle{\frac{1}{2}}}[\omega
,\pi]_{\left\langle {ab}\right\rangle }-{\textstyle{\frac{1}{2}}}\sigma^{e}%
{}_{\left\langle a\right.  }\pi_{\left.  b\right\rangle e}   =0,
\label{eq:e_2_14}%
\end{align}%
\begin{align}
P^{2}{}_{ab}\equiv & \mathrm{curl}(E)_{ab}+2[\dot{u},E]_{\left\langle
{ab}\right\rangle }+\dot{H}_{\left\langle {ab}\right\rangle }+\Theta
H_{ab}-[\omega,H]_{\left\langle {ab}\right\rangle }  \nonumber\\
& -3\sigma_{c\left\langle a\right.  }H_{\left.  b\right\rangle }{}%
^{c}-{\textstyle{\frac{3}{2}}}\omega_{\left\langle a\right.
}q_{\left. b\right\rangle
}-{\textstyle{\frac{1}{2}}}[\sigma,q]_{\left\langle
{ab}\right\rangle }-{\textstyle{\frac{1}{2}}}\mathrm{curl}(\pi)_{ab} =0.
\label{eq:e_2_15}%
\end{align}

The twice contracted Bianchi identities present the conservation of
the total energy momentum tensor, namely
\begin{equation}
\nabla^{b}T_{ab}=\nabla^{b}(R_{ab}-{\textstyle{\frac{1}{2}}}g_{ab}R)=0.
\label{eq:e_2_16}%
\end{equation}
It is split into a timelike and a spacelike momentum constraints:
\begin{equation}
C^{3}\equiv\dot{\rho}+(\rho+p)\Theta+{\mathrm{div}(q)}+2\dot{u}_{a}%
q^{a}+\sigma_{ab}\pi^{ab}=0, \label{eq:e_2_17}%
\end{equation}%
\begin{align}
C^{4}{}_{a}\equiv & (\rho+p)\dot{u}_{a}+\mathrm{D}_{a}p+\dot{q}_{\left\langle
a\right\rangle }+{\textstyle{\frac{4}{3}}}\Theta q_{a}+\sigma_{ab}%
q^{b} \nonumber\\
& -[\omega,q]_{a}+{(\mathrm{div}\pi)_{a}}+\dot{u}^{b}\pi_{ab}=0.
\label{eq:e_2_18}%
\end{align}
They provide the conservation law of energy-momentum, i.\thinspace
e., how matter determines the geometry, and describe the motion of
matter.

\subsection{Kinematic Formulas}

To provide the equations of motion, we use the \textit{Ricci
identities} for the vector field $u_{a}$:
\begin{equation}
2\nabla_{\lbrack a}\nabla_{b]}u_{c}=R_{abcd}u^{d}.\label{eq:e_2_19}%
\end{equation}
We substitute the vector field $u_{a}$ from the kinematic
quantities, using the Einstein equation, and separating out the
orthogonally projected part into trace, symmetric traceless, and
skew symmetric parts. We obtain constraints and propagations for the
kinematic quantities as follows\cite{Trumper1964}:
\begin{equation}
P^{3}\equiv\dot{\Theta}+{\textstyle{\frac{1}{3}}}\Theta^{2}-{\mathrm{div}%
(\dot{u})}-\dot{u}^{a}\dot{u}_{a}-(\omega_{ab}\omega^{ab}-\sigma_{ab}%
\sigma^{ab})+{\textstyle{\frac{1}{2}}}(\rho+3p)=0,\label{eq:e_2_20}%
\end{equation}%
\begin{equation}
P^{4}{}_{a}\equiv\dot{\omega}_{\left\langle a\right\rangle
}+{\textstyle{\frac
{2}{3}}}\Theta\omega_{a}-\sigma_{a}{}^{b}\omega_{b}+{\textstyle{\frac{1}{2}}%
}\mathrm{curl}{(}\dot{u})_{a}=0,\label{eq:e_2_21}%
\end{equation}%
\begin{align}
P^{5}{}_{ab}\equiv & E_{ab}-\mathrm{D}_{\left\langle a\right.
}\dot{u}_{\left. b\right\rangle }-\dot{u}_{\left\langle a\right.
}\dot{u}_{\left. b\right\rangle }+\dot{\sigma}_{\left\langle
{ab}\right\rangle }+\sigma
_{c\left\langle a\right.  }\sigma_{\left.  b\right\rangle }{}^{c}%
+{\textstyle{\frac{2}{3}}}\sigma_{ab}\Theta+\omega_{\left\langle
a\right.
}\omega_{\left.  b\right\rangle }
 \nonumber\\
& -{\textstyle{\frac{1}{2}}}\pi_{ab}%
=0.\label{eq:e_2_22}%
\end{align}
Equation (\ref{eq:e_2_20}), called the Raychaudhuri propagation formula,
is the basic equation of gravitational attraction.\cite{Ellis1999} 
In Eq. (\ref{eq:e_2_21}), the evolution of vorticity is conserved by
the rotation of acceleration. Equation (\ref{eq:e_2_22}) shows that the
gravitoelectric field is propagated in shear, vorticity,
acceleration, and anisotropic stress.

The Ricci identities also provide a set of constraints:
\begin{equation}
C^{5}\equiv{\mathrm{div}(\omega)}-\omega_{a}\dot{u}^{a}=0,\label{eq:e_2_23}%
\end{equation}%
\begin{equation}
C^{6}{}_{a}\equiv{\textstyle{\frac{2}{3}}}\mathrm{D}_{a}\Theta-{(\mathrm{div}%
\sigma)_{a}}+\mathrm{curl}(\omega)_{a}+2[\dot{u},\omega]_{a}-q_{a}%
=0,\label{eq:e_2_24}%
\end{equation}%
\begin{equation}
C^{7}{}_{ab}\equiv
H_{ab}-\mathrm{curl}(\sigma)_{ab}+\mathrm{D}_{\left\langle a\right.
}\omega_{\left.  b\right\rangle }+2\dot{u}_{\left\langle a\right.
}\omega_{\left.  b\right\rangle }=0.\label{eq:e_2_25}%
\end{equation}
Equation (\ref{eq:e_2_23}) presents the divergence of vorticity. Equation 
(\ref{eq:e_2_24}) links the divergence of shear to the rotation of
vorticity. Equation (\ref{eq:e_2_25}) characterizes the gravitomagnetic
field as the distortion of vorticity and the rotation of shear.

\section{Gravitational Waves}

\label{sec4}

We now obtain the temporal evolution of the dynamic propagations ($P^{1,2}%
{}_{ab}$) in a perfect-fluid model ($q^{a}=\pi_{ab}=0$):
\begin{equation}
C^{1}{}_{a}={(\mathrm{div}E)_{a}}-3\omega^{b}H_{ab}-[\sigma,H]_{a}%
-{\textstyle{\frac{1}{3}}}\mathrm{D}_{a}\rho=0,\label{eq:e_4_1}%
\end{equation}%
\begin{equation}
C^{2}{}_{a}={(\mathrm{div}H)_{a}}+3\omega^{b}E_{ab}+[\sigma,E]_{a}+\omega
_{a}(\rho+p)=0,\label{eq:e_4_2}%
\end{equation}%
\begin{align}
P^{1}{}_{ab}= & \mathrm{curl}(H)_{ab}+2[\dot{u},H]_{\left\langle {ab}%
\right\rangle }-\dot{E}_{\left\langle {ab}\right\rangle }-\Theta
E_{ab}+[\omega,E]_{\left\langle {ab}\right\rangle }   \nonumber\\
& +3\sigma_{c\left\langle a\right.  }E_{\left.  b\right\rangle }{}%
^{c}-{\textstyle{\frac{1}{2}}}\sigma_{ab}(\rho+p)  =0,\label{eq:e_4_3}%
\end{align}%
\begin{align}
P^{2}{}_{ab}=&\mathrm{curl}(E)_{ab}+2[\dot{u},E]_{\left\langle {ab}%
\right\rangle }+\dot{H}_{\left\langle {ab}\right\rangle }+\Theta
H_{ab}-[\omega,H]_{\left\langle {ab}\right\rangle
}
\nonumber\\
& -3\sigma_{c\left\langle
a\right.  }H_{\left.  b\right\rangle }{}^{c}=0.\label{eq:e_4_4}%
\end{align}%
To the first order, the evolution of propagation is
\begin{align}
\dot{P}^{1}{}_{ab}= & \mathrm{D}^{2}E_{ab}-\ddot{E}_{\left\langle {ab}%
\right\rangle }-{\textstyle{\frac{3}{2}}}\mathrm{D}_{\left\langle
a\right.
}C^{1}{}_{\left.  b\right\rangle }-{\textstyle{\frac{4}{3}}}\Theta P^{1}%
{}_{ab}+\mathrm{curl}(P^{2}){}_{ab}  \nonumber\\
& -{\textstyle{\frac{4}{3}}}\Theta^{2}E_{ab}-{\textstyle{\frac{7}{3}}}\Theta
\dot{E}_{\left\langle {ab}\right\rangle }-\dot{\Theta}E_{ab}-\Theta
E_{c\left\langle a\right.  }\sigma_{\left.  b\right\rangle
}{}^{c}-\sigma
_{cd}E^{cd}\sigma_{ab}  \nonumber\\
& +E^{cd}\sigma_{ca}\sigma_{bd}-\sigma^{cd}\sigma_{c(a}E_{b)d}+\varepsilon
_{cd(a}\dot{E}_{b)}{}^{c}\omega^{d}+\varepsilon_{cd(a}E_{b)}{}^{c}\dot{\omega
}^{d} \nonumber\\
& +{\textstyle{\frac{4}{3}}}\Theta\lbrack\omega,E]_{\left\langle {ab}%
\right\rangle }+4\Theta\sigma_{c\left\langle a\right.  }E_{\left.
b\right\rangle }{}^{c}+\dot{\varepsilon}_{cd(a}E_{b)}^{c}\omega^{d}%
+3\dot{\sigma}_{c\left\langle a\right.  }E_{\left.  b\right\rangle
}{}^{c} \nonumber\\
& +3\sigma_{c\left\langle a\right.  }\dot{E}_{\left.  b\right\rangle }{}%
^{c}-2\mathrm{curl}([\dot{u},E])_{{ab}}-{\textstyle{\frac{1}{2}}}%
\mathrm{D}_{\left\langle a\right.  }\omega^{c}H_{\left.
b\right\rangle c} \nonumber\\
& -{\textstyle{\frac{3}{2}}}\mathrm{D}_{\left\langle a\right. }[\sigma
,H]_{\left.  b\right\rangle
}+{\textstyle{\frac{8}{3}}}\Theta\lbrack\dot {u},H]_{\left\langle
{ab}\right\rangle }+\mathrm{curl}([\omega,H])_{{ab}} \nonumber\\
& +3\mathrm{curl}\left(  {\sigma_{c\left\langle a\right.  }H_{\left.
b\right\rangle }{}^{c}}\right)  -\sigma_{e}{}^{c}\varepsilon_{cd(a}%
\mathrm{D}^{e}H_{b)}{}^{d}+2\varepsilon_{cd(a}\dot{H}_{b)}{}^{c}\dot{u}^{d}
\nonumber\\
& +2\varepsilon_{cd(a}H_{b)}{}^{c}\ddot{u}^{d}+2\dot{\varepsilon}_{cd(a}H_{b)}%
{}^{c}\dot{u}^{d}-{\textstyle{\frac{1}{2}}}\sigma_{ab}(\dot{\rho}+\dot{p}) \nonumber\\
& -{\textstyle{\frac{1}{3}}}\Theta\sigma_{ab}(\rho+p)  =0.\label{eq:e_4_5}%
\end{align}
We neglect products of kinematic quantities with respect to the
undisturbed metrics (unexpansive static spacetime). We can also
prevent the perturbations that are merely associated with coordinate
transformation, since they have no physical significance. In free
space, we get
\begin{equation}
\dot{P}^{1}{}_{ab}=\mathrm{D}^{2}E_{ab}-\ddot{E}_{\left\langle {ab}%
\right\rangle }-{\textstyle{\frac{3}{2}}}\mathrm{D}_{\left\langle
a\right.
}C^{1}{}_{\left.  b\right\rangle }-{\textstyle{\frac{4}{3}}}\Theta P^{1}%
{}_{ab}+\mathrm{curl}(P^{2}){}_{ab}=0.\label{eq:e_4_6}%
\end{equation}
To be consistent with Eqs. (\ref{eq:e_4_1})--(\ref{eq:e_4_4}), $\mathrm{D}%
^{2}E_{ab}-\ddot{E}_{\left\langle {ab}\right\rangle }$ has to
vanish.
Similarly, the evolution of $P^{2}{}_{ab}$ shows that $\mathrm{D}^{2}%
H_{ab}-\ddot{H}_{\left\langle {ab}\right\rangle }=0$. The evolutions
reflect that the divergenceless and nonvanishing rotation of the
Weyl fields are necessary conditions for gravitational waves:
\begin{equation}%
\begin{array}
[c]{cc}%
{(\mathrm{div}E)_{a}\mathrm{\ =}(\mathrm{div}H)_{a}=0,} & {~~~~~~~~\mathrm{curl}%
(E){}_{ab}\neq0\neq\mathrm{curl}(H){}_{ab}.}%
\end{array}
\label{eq:e_4_7}%
\end{equation}
Indeed, the rotation of the Weyl fields characterizes the wave
solutions. The gravitomagnetic field is explicitly important to
describe the gravitational waves, and is comparable with the Maxwell
fields.

We use Eq. (\ref{eq:e_3_26}) to provide two more constraints:
\begin{equation}
C^{8}{}_{abc}\equiv\mathrm{D}_{a}E_{bc}-\mathrm{D}_{\left\langle
a\right.
}E_{\left.  {bc}\right\rangle }-{\textstyle{\frac{3}{5}}}\mathrm{D}%
^{d}E_{d\left\langle a\right.  }h_{\left.  b\right\rangle
c}+{\textstyle{\frac
{2}{3}}}\varepsilon_{dc(a}\mathrm{curl}(E)_{b)}{}^{d}=0,\label{eq:e_4_8}%
\end{equation}%
\begin{equation}
C^{9}{}_{abc}\equiv\mathrm{D}_{a}H_{bc}-\mathrm{D}_{\left\langle
a\right.
}H_{\left.  {bc}\right\rangle }-{\textstyle{\frac{3}{5}}}\mathrm{D}%
^{d}H_{d\left\langle a\right.  }h_{\left.  b\right\rangle
c}+{\textstyle{\frac
{2}{3}}}\varepsilon_{dc(a}\mathrm{curl}(H)_{b)}{}^{d}=0.\label{eq:e_4_9}%
\end{equation}
To first order, divergence of Eq. (\ref{eq:e_4_8}) is
\begin{align}
\mathrm{D}^{a}C^{8}{}_{abc}= & \mathrm{D}^{2}E_{bc}-\mathrm{D}^{a}\mathrm{D}%
_{\left\langle a\right.  }E_{\left.  {bc}\right\rangle
}-{\textstyle{\frac
{3}{5}}}\mathrm{D}^{a}\mathrm{D}^{d}E_{d\left\langle a\right.
}h_{\left.
b\right\rangle c}  \nonumber\\
& +{\textstyle{\frac{1}{3}}}\mathrm{D}^{a}\varepsilon_{dca}\mathrm{curl}%
(E)_{b}{}^{d}+{\textstyle{\frac{1}{3}}}\varepsilon_{dcb}\mathrm{D}%
^{a}\mathrm{curl}(E)_{a}{}^{d} =0.
\end{align}
On substituting Eq. (\ref{eq:e_4_4}), it becomes
\begin{align}
\mathrm{D}^{a}C^{8}{}_{abc}= & \mathrm{D}^{2}E_{bc}-\mathrm{D}^{a}\mathrm{D}%
_{\left\langle a\right.  }E_{\left.  {bc}\right\rangle
}-{\textstyle{\frac
{3}{5}}}\mathrm{D}^{a}\mathrm{D}^{d}E_{d\left\langle a\right.
}h_{\left. b\right\rangle
c}-{\textstyle{\frac{3}{5}}}\mathrm{D}^{a}C^{1}{}_{\left\langle
a\right.  }h_{\left.  b\right\rangle c} \nonumber\\
& +{\textstyle{\frac{2}{3}}}\mathrm{D}^{a}\varepsilon^{d}{}_{c(a}P^{2}{}%
_{b)d}-{\textstyle{\frac{9}{5}}}\mathrm{D}^{a}\omega^{d}H_{d\left\langle
a\right.  }h_{\left.  b\right\rangle c}-{\textstyle{\frac{3}{5}}}%
\mathrm{D}^{a}[\sigma,H]_{\left\langle a\right.  }h_{\left.
b\right\rangle c}
\nonumber\\
& -{\textstyle{\frac{1}{5}}}\mathrm{D}^{a}\mathrm{D}_{\left\langle
a\right.
}\rho h_{\left.  b\right\rangle c}-{\textstyle{\frac{2}{3}}}\mathrm{D}%
^{a}\varepsilon^{d}{}_{c(a}\dot{H}_{\left\langle {b)d}\right\rangle
}-{\textstyle{\frac{4}{3}}}\mathrm{D}^{a}\varepsilon^{d}{}_{c(a}[\dot
{u},E]_{\left\langle {b)d}\right\rangle } \nonumber\\
& -\Theta{\textstyle{\frac{2}{3}}}\mathrm{D}^{a}\varepsilon^{d}{}_{c(a}%
H_{b)d}+{\textstyle{\frac{2}{3}}}\mathrm{D}^{a}\varepsilon^{d}{}_{c(a}%
[\omega,H]_{\left\langle {b)d}\right\rangle } \nonumber\\
& +2\mathrm{D}^{a}\varepsilon^{d}{}_{c(a}\sigma_{c\left\langle
{b)}\right.
}H_{\left.  d\right\rangle }{}^{c} =0.\label{eq:e_4_11}%
\end{align}
We abandon products of kinematic quantities in the undisturbed
metrics:
\begin{align}
\mathrm{D}^{a}C^{8}{}_{abc}= & \mathrm{D}^{2}E_{bc}-\mathrm{D}^{a}\mathrm{D}%
_{\left\langle a\right.  }E_{\left.  {bc}\right\rangle
}-{\textstyle{\frac
{3}{5}}}\mathrm{D}^{a}\mathrm{D}^{d}E_{d\left\langle a\right.
}h_{\left. b\right\rangle
c}-{\textstyle{\frac{3}{5}}}\mathrm{D}^{a}C^{1}{}_{\left\langle
a\right.  }h_{\left.  b\right\rangle c}  \nonumber\\
& +{\textstyle{\frac{2}{3}}}\mathrm{D}^{a}\varepsilon^{d}{}_{c(a}P^{2}{}%
_{b)d}-{\textstyle{\frac{2}{3}}}\mathrm{curl}(\dot{H})_{ab} 
=0.\label{eq:e_4_12}%
\end{align}
To linearized order, we get $\mathrm{curl}(S_{ab})^{\cdot}=\mathrm{curl}%
\dot{S}_{ab}$. Using the later point and the evolution of Eq. (\ref{eq:e_4_3}%
), we obtain:
\begin{align}
\mathrm{D}^{a}C^{8}{}_{abc}= & \mathrm{D}^{2}E_{bc}-\mathrm{D}^{a}\mathrm{D}%
_{\left\langle a\right.  }E_{\left.  {bc}\right\rangle
}-{\textstyle{\frac
{3}{5}}}\mathrm{D}^{a}\mathrm{D}^{d}E_{d\left\langle a\right.
}h_{\left. b\right\rangle
c}-{\textstyle{\frac{2}{3}}}\ddot{E}_{\left\langle
{ab}\right\rangle } \nonumber\\
& -{\textstyle{\frac{3}{5}}}\mathrm{D}_{\left\langle a\right.
}C^{1}{}_{\left.
b\right\rangle }+{\textstyle{\frac{2}{3}}}\mathrm{D}^{a}\varepsilon^{d}%
{}_{c(a}P^{2}{}_{b)d}-{\textstyle{\frac{2}{3}}}\dot{P}^{1}{}_{ab}
=0.\label{eq:e_4_13}%
\end{align}
The result can be compared to the wave solution (\ref{eq:e_4_6}).
Without the distortion parts, it is inconsistent with a generic
description of wave. Distortion of the gravitoelectric field
($\mathrm{D}_{\left\langle a\right. }E_{\left.  {bc}\right\rangle
}$) must not vanish to provide the wave solution. We also obtain 
similar condition for the gravitomagnetic field. In free space, the
divergence of the Weyl fields, determined by the matter, must be
free. The temporal evolution decides that the rotation of the Weyl
fields must be non-zero. Now, the distortion provides another
condition to characterize the evolution of the Weyl fields:
\begin{equation}
\mathrm{D}_{\left\langle a\right.  }E_{\left.  {bc}\right\rangle
}\neq 0\neq\mathrm{D}_{\left\langle a\right.  }H_{\left.
{bc}\right\rangle
}.\label{eq:e_4_14}%
\end{equation}
The existence of rotation and distortion is necessary condition to
maintain the wave solutions.

\section{Newtonian and Anti-Newtonian Fields}

\label{sec5}

We can associate a Newtonian model with purely gravitoelectric
($H_{ab}=0$). Without the gravitomagnetism, the nonlocal nature of
the Newtonian force cannot be retrieved from relativistic models. It
also excludes gravitational waves. The Newtonian model is a limited
model to show the characteristics of the gravitoelectric. We can
also consider an anti-Newtonian model; a model with purely
gravitomagnetic ($E_{ab}=0$). The anti-Newtonian model obstructs
sounding solutions. In Ref.~\onlinecite{Trumper1965}, it has been proven that
the anti-Newtonian model shall include either shear or vorticity.

\subsection{Newtonian Model}

Let us consider the Newtonian model ($H_{ab}=0$) in an irrotational
static
spacetime ($\omega_{a}=\dot{u}_{a}=0$) and a perfect-fluid model ($q^{a}%
=\pi_{ab}=0$). The constraints and propagations shall be
\begin{equation}%
\begin{array}
[c]{cc}%
{C^{1}{}_{a}=(\mathrm{div}E)_{a}-{\textstyle{\frac{1}{3}}}\mathrm{D}_{a}%
\rho=0,} & {~~~~~~~~C^{2}{}_{a}=[\sigma,E]_{a}=0,}%
\end{array}
\label{eq:e_5_1}%
\end{equation}%
\begin{equation}%
\begin{array}
[c]{c}%
{P^{1}{}_{ab}=-\dot{E}_{\left\langle {ab}\right\rangle }-\Theta E_{ab}%
+3\sigma_{c\left\langle a\right.  }E_{\left.  b\right\rangle }{}%
^{c}-{\textstyle{\frac{1}{2}}}\sigma_{ab}(\rho+p)=0,}\\
\\
{P^{2}{}_{ab}=\mathrm{curl}(E)_{ab}=0,}%
\end{array}
\label{eq:e_5_3}%
\end{equation}%
\begin{equation}%
\begin{array}
[c]{cc}%
{C^{6}{}_{a}={\textstyle{\frac{2}{3}}}\mathrm{D}_{a}\Theta-(\mathrm{div}%
\sigma)_{a}=0,} & {~~~~~~~~C^{7}{}_{ab}=-\mathrm{curl}(\sigma)_{ab}=0.}%
\end{array}
\label{eq:e_5_2}%
\end{equation}
To first order, divergence and evolution of Eq.
(\ref{eq:e_5_3}b) are
\begin{align}
\mathrm{D}^{b}P^{2}{}_{ab}  = &{\textstyle{\frac{1}{2}}}\varepsilon
_{abc}\mathrm{D}^{b}\mathrm{(D}_{d}E^{cd})+{\textstyle{\frac{1}{3}}}%
\Theta\lbrack\sigma,E]_{a}-\sigma_{ab}[\sigma,E]^{b}\nonumber\\
= & {\textstyle{\frac{1}{2}}}\mathrm{curl}(C^{1}){}_{a}+{\textstyle{\frac
{1}{3}}}\Theta C^{2}{}_{a}-\sigma_{a}{}^{b}C^{2}{}_{b}+{\textstyle{\frac{1}%
{3}}}\omega{}_{a}\dot{\rho}, \label{eq:e_5_4}%
\end{align}%
\begin{align}
\dot{P}^{2}{}_{ab}   = & -{\textstyle{\frac{1}{3}}}\Theta{\mathrm{curl}%
(E)_{ab}}-\sigma_{e}{}^{c}\varepsilon_{cd(a}\mathrm{D}^{e}E_{b)}{}%
^{d}+{\mathrm{curl}(\dot{E})}_{{ab}}\nonumber\\
 = & -{\textstyle{\frac{3}{2}}}\varepsilon^{cd}{}_{(a}\sigma_{b)c}C^{1}{}%
_{d}-{\textstyle{\frac{4}{3}}}\Theta P^{2}{}_{ab}+{\textstyle{\frac{3}{2}}%
}\varepsilon^{c}{}_{d(a}C^{6}{}_{c}E_{b)}{}^{d}\nonumber\\
&  -{\textstyle{\frac{1}{2}}}(\rho+p)C^{7}{}_{ab}-\mathrm{curl}(P^{1}){}%
_{ab}+3\mathrm{curl(}\sigma_{c\left\langle a\right.  }E_{\left.
b\right\rangle }{}^{c}). \label{eq:e_5_5}%
\end{align}
The last parameter (${\textstyle{\frac{1}{3}}}\omega_{a}\dot{\rho}$)
in Eq. (\ref{eq:e_5_4}) vanishes because of irrotational condition.
Equation (\ref{eq:e_5_4}) then conserves the constraints. Equation 
(\ref{eq:e_5_5}) must be consistent with Eqs. (\ref{eq:e_5_1}) and
(\ref{eq:e_5_2}). Thus, the last parameters in Eq. (\ref{eq:e_5_5})
has to vanish:
\begin{equation}
\mathrm{curl}(\sigma_{c\left\langle a\right.  }E_{\left.  b\right\rangle }%
{}^{c})=0. \label{eq:e_5_6}%
\end{equation}
It is a necessary condition for the consistent evolution of
propagation. This condition is satisfied with irrotational product
of gravitoelectric and shear, but it is a complete contrast to Eq.
(\ref{eq:e_5_1}b). Thus, the Newtonian model is generally
inconsistent with generic relativistic models. Moreover, the
temporal evolution of propagation shows no wave solutions.

\subsubsection{Newtonian Limit}

The Newtonian model obstructs wave solution, due to the
instantaneous interaction. Following 
Refs.~\onlinecite{Heckmann1955,Heckmann1956,Heckmann1959}, we consider a model
whose action propagates at infinite speed ($c\rightarrow \infty$).
This is compatible with $\mathop{\lim }\limits_{c\rightarrow\infty
}E_{ab}=E_{ab}(t)|_{\infty}$, where $E_{ab}(t)|_{\infty}$ is an
arbitrary function of time.

We define the Newtonian potential as
\begin{equation}
E_{ab}\equiv\mathrm{D}_{\left\langle a\right.  }\mathrm{D}_{\left.
b\right\rangle }\Phi=\mathrm{D}_{a}\mathrm{D}_{b}\Phi-{\textstyle{\frac{1}{3}%
}}h_{ab}\mathrm{D}^{2}\Phi.\label{eq:e_5_7}%
\end{equation}
On substituting into Eq. (\ref{eq:e_5_1}a), we get
\begin{equation}
C^{1}{}_{a}=\mathrm{D}_{a}\mathrm{D}^{2}\Phi-{\textstyle{\frac{1}{3}}%
}\mathrm{D}^{b}h_{ab}\mathrm{D}^{2}\Phi-{\textstyle{\frac{1}{3}}}%
\mathrm{D}_{a}\rho=0.\label{eq:e_5_8}%
\end{equation}
In a spatial infinity, we obtain the Poisson equation of the
Newtonian
potential:%
\begin{equation}
C^{1}\equiv\mathrm{D}^{2}\Phi-{\textstyle{\frac{1}{2}}}\rho
=0.\label{eq:e_5_8_1}%
\end{equation}
Equation (\ref{eq:e_5_1}a) generalizes the gravitoelectric as the
Newtonian force in the gradient of the relativistic energy density.

Moreover, Eq. (\ref{eq:e_5_3}a) gives
\begin{align}
P^{1}{}_{ab}=-\mathrm{D}_{a}\mathrm{D}_{b}\dot{\Phi}+{\textstyle{\frac{1}{3}}%
}h_{ab}\mathrm{D}^{2}\dot{\Phi}-\Theta\mathrm{D}_{a}\mathrm{D}_{b}%
\Phi+{\textstyle{\frac{1}{3}}}(\dot{h}_{ab}+\Theta
h_{ab}\mathrm{)D}^{2}\Phi &
\nonumber\\
+3\sigma_{c\left\langle a\right.  }\mathrm{D}_{\left. b\right\rangle
}\mathrm{D}^{c}\Phi-\sigma_{c\left\langle a\right. }h_{\left.
b\right\rangle
}{}^{\mathrm{c}}\mathrm{D}^{2}\Phi-{\textstyle{\frac{1}{2}}}\sigma_{ab}%
(\rho+p)  &  =0. \label{eq:e_5_9}%
\end{align}
In the Newtonian theory, we could not find the temporal evolution of
the Newtonian potential.

\subsubsection{Acceleration Potential}

In an irrotational spacetime, Eq. (\ref{eq:e_2_21}) becomes
\begin{equation}
P^{4}{}_{a}={\textstyle{\frac{1}{2}}}\mathrm{curl}(\dot{u}){}_{a}=0.
\label{eq:e_5_10}%
\end{equation}
It introduces a scalar potential:
\begin{equation}
\dot{u}_{a}=\mathrm{D}_{a}\Phi, \label{eq:e_5_11}%
\end{equation}
where $\Phi$ is the acceleration potential. This scalar potential
corresponds to the Newtonian potential. In the irrotational
Newtonian model, the linearized acceleration is characterized as the
acceleration potential.

\subsection{Anti-Newtonian Model}

Let us consider the anti-Newtonian model ($E_{ab}=0$) in a shearless
static
spacetime ($\omega_{a}=\dot{u}_{a}=0$) and a perfect-fluid model ($q^{a}%
=\pi_{ab}=0$). The constraints and propagations shall be
\begin{equation}%
\begin{array}
[c]{cc}%
{C^{1}{}_{a}=-3\omega^{b}H_{ab}-{\textstyle{\frac{1}{3}}}\mathrm{D}_{a}%
\rho=0,} & {~~~~~~~~C^{2}{}_{a}=(\mathrm{div}H)_{a}+\omega_{a}(\rho+p)=0,}%
\end{array}
\label{eq:e_5_13}%
\end{equation}%
\begin{equation}%
\begin{array}
[c]{cc}%
{P^{1}{}_{ab}=\mathrm{curl}(H)_{ab}=0,} &
{~~~~~~~~P^{2}{}_{ab}=\dot{H}_{\left\langle {ab}\right\rangle }+\Theta
H_{ab}-[\omega,H]_{\left\langle {ab}\right\rangle
}=0,}%
\end{array}
\label{eq:e_5_14}%
\end{equation}%
\begin{equation}%
\begin{array}
[c]{cc}%
{C^{6}{}_{a}={\textstyle{\frac{2}{3}}}\mathrm{D}_{a}\Theta+\mathrm{curl}%
(\omega)_{a}=0,} & {~~~~~~~~C^{7}{}_{ab}=H_{ab}+\mathrm{D}_{\left\langle
a\right.
}\omega_{\left.  b\right\rangle }+2\dot{u}_{\left\langle a\right.  }%
\omega_{\left.  b\right\rangle }=0.}%
\end{array}
\label{eq:e_5_15}%
\end{equation}
To linearized order, divergence and evolution of Eq.
(\ref{eq:e_5_14}a) are
\begin{align}
\mathrm{D}^{b}P^{1}{}_{ab}  = & {\textstyle{\frac{1}{2}}}\varepsilon
_{abc}\mathrm{D}^{b}\mathrm{(D}_{d}H^{cd})\nonumber\\
= & {\textstyle{\frac{1}{2}}}\varepsilon_{ab}{}^{c}\mathrm{D}^{b}C^{2}{}%
_{c}-{\textstyle{\frac{1}{2}}}(\rho+p)C^{6}{}_{a}+{\textstyle{\frac{1}{3}}%
}(\rho+p)\mathrm{D}_{a}\Theta, \label{eq:e_5_16}%
\end{align}%
\begin{align}
\dot{P}^{1}{}_{ab}   = &-{\textstyle{\frac{1}{3}}}\Theta\mathrm{curl}%
{(H)_{ab}}+{\mathrm{curl}(\dot{H})_{ab}}\nonumber\\
= &  -{\textstyle{\frac{4}{3}}}\Theta P^{1}{}_{ab}+\mathrm{curl}(P^{2}){}%
_{ab}+\mathrm{curl}([\omega,H])_{\left\langle {ab}\right\rangle }.
\label{eq:e_5_17}%
\end{align}
Equation (\ref{eq:e_5_16}) is consistent only in the spacetime being free
from either the gravitational mass and pressure or the gradient of
expansion. According to Eqs. (\ref{eq:e_5_13}) and
(\ref{eq:e_5_15}), the last term in Eq. (\ref{eq:e_5_17}) has to
vanish:
\begin{equation}
\mathrm{curl}([\omega,H])_{\left\langle {ab}\right\rangle }=0.
\label{eq:e_5_18}%
\end{equation}
It is a necessary condition for the consistent evolution of
propagation. This condition is satisfied with irrotational vorticity
products of gravitomagnetic, but it is not consistent with Eq.
(\ref{eq:e_5_15}b):
\begin{align}
\varepsilon^{c}{}_{d(a}C^{7}{}_{b)c}\omega^{d}-{\textstyle{\frac{1}{4}}}%
\omega_{b}C^{6}{}_{a}-{\textstyle{\frac{1}{4}}}\omega_{a}C^{6}{}%
_{b}-{\textstyle{\frac{1}{4}}}\mathrm{D}_{b}[\omega,\omega]_{a}%
-{\textstyle{\frac{1}{4}}}\mathrm{D}_{a}[\omega,\omega]_{b}  & \nonumber\\
+{\textstyle{\frac{1}{6}}}\omega_{b}\mathrm{D}_{a}\Theta+{\textstyle{\frac
{1}{6}}}\omega_{a}\mathrm{D}_{b}\Theta-\varepsilon^{c}{}_{da}\dot
{u}_{\left\langle b\right.  }\omega_{\left.  c\right\rangle }\omega
^{d}-\varepsilon^{c}{}_{db}\dot{u}_{\left\langle a\right.
}\omega_{\left.
c\right\rangle }\omega^{d}  &  =0. \label{eq:e_5_23}%
\end{align}
Thus, the anti-Newtonian model is generally inconsistent with
relativistic models. Furthermore, there is not a possibility of
gravitational waves.

\subsubsection{Vorticity Potential}

In an unexpansive spacetime, Eq. (\ref{eq:e_5_15}a) takes the
following form:
\begin{equation}
C^{6}{}_{a}=\mathrm{curl}{(\omega)_{a}}=0.\label{eq:e_5_19}%
\end{equation}
It defines a vorticity scalar potential $\Psi$ as%
\begin{equation}
\omega_{a}=\mathrm{D}_{a}\Psi.\label{eq:e_5_20}%
\end{equation}
In the unexpansive anti-Newtonian model, the linearized vorticity is
characterized as the vorticity potential.

\subsubsection{Anti-Newtonian Limit}

We may consider a gravitomagnetic model whose action propagates at
infinite speed. Let us define the anti-Newtonian potential as
\begin{equation}
H_{ab}\equiv\mathrm{D}_{\left\langle a\right.  }\mathrm{D}_{\left.
b\right\rangle }\Psi=\mathrm{D}_{a}\mathrm{D}_{b}\Psi-{\textstyle{\frac{1}{3}%
}}h_{ab}\mathrm{D}^{2}\Psi. \label{eq:e_5_21}%
\end{equation}
We substitute Eqs. (\ref{eq:e_5_20}) and (\ref{eq:e_5_21}) into Eq.
(\ref{eq:e_5_13}b):
\begin{equation}
C^{2}{}_{a}=\mathrm{D}_{a}\mathrm{D}^{2}\Psi-{\textstyle{\frac{1}{3}}%
}\mathrm{D}^{b}h_{ab}\mathrm{D}^{2}\Psi+\mathrm{D}_{a}\Psi(\rho+p)=0.
\label{eq:e_5_22}%
\end{equation}
In a spatial infinity, we derive the Helmholtz equation:%
\begin{equation}
C^{2}\equiv\mathrm{D}^{2}\Psi+{\textstyle{\frac{3}{2}}}(\rho+p)\Psi=0.
\label{eq:e_5_22_1}%
\end{equation}
Eq. (\ref{eq:e_5_13}b) associates the gravitomagnetic with the
angular momentum $\omega_{a}(\rho+p)$.

\section{Conclusion}

\label{sec6}

The Weyl curvature tensor describes the nonlocal long-range
interactions as enabling gravitational act at a distance (tidal
forces and gravitational waves). The gravitoelectric field is
described as the relativistic generalization of the tidal
(Newtonian) force. However, the gravitomagnetic (anti-Newtonian)
force has no Newtonian analogue. We have no expression similar to
$\dot{E}_{ab}$ in the Newtonian theory. This difference arises from
the instantaneous action in the Newtonian theory, which excludes a
sounding solution. In Sec.~\ref{sec4}, the rotation and distortion of
the Weyl fields characterize the gravitational wave. The
gravitomagnetism is necessary to maintain the gravitational wave. In
relativistic models, the Newtonian force is also inconsistent
without the magnetic part of the Weyl curvature.

\section*{Acknowledgment}

I have been supported by a Marie Curie Action under FP6 from the EU contract MRTN-CT-2004-005104 during my stay at the University of Craiova. I am also indebted to a referee for valuable comments.


\begin{thebibliography}{99}                                                                                               %


\bibitem {Jordean1960}P.~Jordean, J.~Ehlers and W.~Kundt, \textit{Ahb. Akad
Wiss. Mainz}, No. \textbf{7} (1960).

\bibitem {Ciufolini1995}I.~Ciufolini and J.\,A.~Wheeler,
\textit{Gravitation and Inertia} (Princeton Univ. Press,
1995).

\bibitem {Hawking1973}S.\,W.~Hawking and G.\,F.\,R.~Ellis, \textit{The Large
Scale Structure of Space-time} (Cambrige University Press, 1973).

\bibitem {Misner1973}C.\,M.~Misner, K.S. Thorne and J. A. Wheeler,
\textit{ Gravitation} (Freeman, 1973).

\bibitem {Thorne1980}K.\,S.~Thorne, \textit{Rev. Mod. Phys.}
\textbf{52}, 299 (1980).

\bibitem {Pirani1957}F.\,A.\,E.~Pirani, \textit{Phys. Rev.} \textbf{105}, 1089 (1957).

\bibitem {Pirani1962a}F.\,A.\,E.~Pirani, in \textit{Recent
Development in General Relativity} (Pergamon Press, 1962).

\bibitem {Pirani1962b}F.\,A.\,E.~Pirani,
in \textit{Gravitation: An introduction to Current Research},
L.~Witten (ed.) (John Wiley \& Sons, 1962).

\bibitem {Kofman1995}L.~Kofman and D.~Pogosayn, \textit{Astrophys. J.}
\textbf{442}, 30 (1995).

\bibitem {Ellis1973}G.\,F.\,R.~Ellis, in \textit{Carg\'{e}se Lectures in
physics}, Vol. 6, ed. E.~Schatzmann (Gordon and Breach, 1973).

\bibitem {Wiltshire2008}D.\,L.~Wiltshire, \textit{Phys. Rev. D}
\textbf{78}, 084032 (2008).

\bibitem {Will2006}C.\,M.~Will, \textit{Living Rev. Rel.} \textbf{9}, 3 (2006).

\bibitem {Hamilton1992}W.\,O.~Hamilton, in \textit{Proc. 6th Marcel
Grossmann Meeting on General Relativity}, eds. H.~Sato and T.~Nakamura
(World Scienetific, 1992).

\bibitem {Bender1989}P.\,L.~Bender, J.\,E.~Faller, D.~Hils
and R.\,T.~Stebbins, in \textit{12th Int. Conf. on General
Relativity and Gravitation}, Boulder, CO (July 1989).

\bibitem {Faller1989}J.\,E.~Faller, P.\,L.~Bender,
J.\,L.~Hall, D.~Hils, R.\,T.~Stebbins and M.\,A.~Vincent,
\textit{Adv. Space Res.} \textbf{9}, 11 (1989).

\bibitem {Fritschel1998}P.~Fritschel, in \textit{Proc. of 2nd Edoardo Amaldi
Conference on Gravitational Waves}, E.~Coccia, G.~Veneziano, and
G.~Pizzella (eds.), (CERN, Switzerland, 1997), \textit{Edoardo
Amaldi Foundation Series} (World Scientific,
1998), pp. 74--85.

\bibitem {Brillet1998}A.~Brillet, in \textit{Proc. of 2nd Edoardo Amaldi
Conference on Gravitational Waves}, E.~Coccia, G.~Veneziano, and
G.~Pizzella (eds.), (CERN, Switzerland, 1997), \textit{Edoardo
Amaldi Foundation Series} (World Scientific,
1998), pp. 86--96.

\bibitem {Bardeen1975}J.\,M.~Bardeen and J.\,A.~Petterson,
\textit{Astrophys. J. Lett.} \textbf{195}, L65 (1975).

\bibitem {Thorne1986}Eds. K.\,S.~Thorne, R.\,H.~Price, and
D.\,A.~MacDonald, \textit{Black Holes, the Membrane Paradiagm}
(Yale University Press, 1986).

\bibitem {Ciufolini2002}I.~Ciufolini, in \textit{Proc. of 22nd Physics in
Collision Conference}, Stanford, California, 2002.

\bibitem {Iorio2005}L.~Iorio, \textit{New Astron.} \textbf{10}, 603 (2005).

\bibitem {Ciufolini2004}I.~Ciufolini and E.\,C.~Pavlis, \textit{Nature}
\textbf{431}, 958 (2004).

\bibitem {Ciufolini2006}I.~Ciufolini, E.\,C.~Pavlis, R.~Peron, \textit{New
Astron.} \textbf{11}, 527 (2006).

\bibitem {Ciufolini2008}I.~Ciufolini, et al, in \textit{Proc. of the First
International School of Astrophysical Relativity \textquotedblright
John Archibald Wheeler\textquotedblright}, Erice, Italy, 2006, eds. I.
Ciufolini and R. Matzner (Springer, 2008).

\bibitem {Raychaudhuri1955}A.~Raychaudhuri, \textit{Phys. Rev.} \textbf{98},
1123 (1955).

\bibitem {Raychaudhuri1957}A.~Raychaudhuri, \textit{Z. Astrophys.}
\textbf{43}, 161 (1957).

\bibitem {Raychaudhuri1979}A.~Raychaudhuri, \textit{Theoretical Cosmology}
(Clarendon Press, 1979).

\bibitem {Heckmann1955}O.~Heckmann and E.~Sch\"{u}cking, \textit{Z. Astrophys.} \textbf{38}, 95 (1955).

\bibitem {Heckmann1956}O.~Heckmann and E.~Sch\"{u}cking, \textit{Z. Astrophys.} \textbf{40}, 81 (1956).

\bibitem {Heckmann1959}O.~Heckmann and E.~Sch\"{u}cking, \textit{Handbuch der
Physik} Vol. 53 (Springer-Verlag, 1959).

\bibitem {Ehlers1993a}J.~Ehlers, \textit{Abh. Akad. Wiss. Lit. Mainz. Nat. Kl} \textbf{11}, 793 (1961).

\bibitem {Ehlers1993b}J.~Ehlers, \textit{Gen. Rel.
Grav.} \textbf{25}, 1225 (1993) [translation of Ref.~\onlinecite{Ehlers1993a}].

\bibitem {Ellis1971}G.\,F.\,R.~Ellis, in \textit{General
Relativity and Cosmology}, ed. R. K.~Sachs (Academic Press, 1971).

\bibitem {Ellis1989}G.\,F.\,R.~Ellis and M.~Bruni, \textit{Phys. Rev. D}
\textbf{40} 1804 (1989).

\bibitem {King1973}A.\,R.~King and G.\,F.\,R.~Ellis, \textit{Commun. Math.
Phys.} \textbf{31}, 209 (1973).

\bibitem {Cattaneo1959}C.~Cattaneo, \textit{Ann. Mat. Pura Appl.} \textbf{48},
86 (1959).

\bibitem {Ryan1975}M.\,P.~Ryan and L.\,C.~Shepley, \textit{Homogeneous
Relativistic Cosmologies} (Princeton Univ. Press, 1975).

\bibitem {Godel1949}K.~G\"{o}del, \textit{Rev. Mod. Phys.} \textbf{21}, 447 (1949).

\bibitem {Godel1950}K.~G\"{o}del, in \textit{Proc. Int. Congress of Math.}
\textbf{1}, 81 (1950).

\bibitem {Kundt1960}W.~Kundt and M.~Tr\"{u}mper, \textit{Akad. Wiss. (Mainz)
Abhandl. Math. Nat. Kl.} \textbf{12}, 970 (1960).

\bibitem {Kundt1962}W.~Kundt and M.~Tr\"{u}mper, \textit{Akad. Wiss. (Mainz)
Abhandl. Math. Nat. Kl.} \textbf{12}, 1 (1962).

\bibitem {Trumper1964}M.~Tr\"{u}mper, \textit{Contribution to Actual Problems
in General Relativity}, preprint 1964.

\bibitem {Trumper1965}M.~Tr\"{u}mper, \textit{J. Math. Phys.} \textbf{6}, 584 (1965).

\bibitem {Trumper1967}M.~Tr\"{u}mper, \textit{Z. Astrophys.}
\textbf{66}, 215 (1967).

\bibitem {Ellis1999}G.\,F.\,R.~Ellis, \textit{Class. Quantum
Grav.} \textbf{16}, A37 (1999).

\end{thebibliography}
\end{document}